\documentclass[english,russian,a4paper,12pt]{article}
\usepackage[T2A]{fontenc}
\usepackage[cp1251]{inputenc}
\usepackage{babel}
\usepackage{amsmath}
\usepackage{graphicx}
\usepackage{amssymb}
\usepackage{epsf}
\usepackage{pazh}
\usepackage[labelsep=period,hang]{caption} 

\tightenlines

\newcommand{\WID}[2]{#1_{\mathrm{#2}}}

\newcommand{\DERS}[3]{\left(\frac{\partial #1}{\partial #2}\right)_{\!\!#3}}
\newcommand{\DERPT}[1]{\left(\frac{\partial #1}{\partial
T}\right)_{\!\!\mathrm{pt}}}

\sloppypar

\begin{document}
\baselineskip 21pt
\def\figurename{Fig.}

\title{\bf SOME PROPERTIES OF CONVECTION IN HYBRID STARS}

\author{\bf \hspace{-1.3cm}\
A.V. Yudin\affilmark{1,2*}, M. Hempel\affilmark{3}, D.K.
Nadyozhin\affilmark{1,4}, T.L. Razinkova\affilmark{1}}

\affil{ $^1${\it Institute for Theoretical and Experimental Physics, Moscow}\\
$^2${\it Novosibirsk State University}\\
$^3${\it Department of Physics, University of Basel, Basel, Switzerland }\\
$^4${\it SRC ``Kurchatov Institute'', Moscow}}

\vspace{2mm}

\sloppypar \vspace{2mm} \noindent
 {\bf Abstract.} It is shown that the unusual thermodynamic properties of matter within the region of
two-phase coexistence in hybrid stars result in a change of the
standard condition for beginning of convection. In particular, the
thermal flux transported by convection may be directed towards the
stellar center. We discuss favorable circumstances leading to such
an effect of ``inverse convection'' and its possible influence on
the thermal evolution of hybrid stars.


\vfill
\noindent\rule{8cm}{1pt}\\
{$^*$ e-mail $<$yudin@itep.ru$>$}

\clearpage

\section*{INTRODUCTION}
\noindent Neutron stars, together with stellar mass black holes,
develop from collapsing cores of massive stars at the final stages
of their evolution. The birth of a neutron star most likely
manifests itself as a supernova explosion. The major part of
matter in neutron stars proves to be in an extreme state with a
density exceeding the nuclear one $\WID{\rho}{n}\approx 2.6\times
10^{14}~\mbox{g/cm}^3$.

The possibility of phase transitions in nuclear matter was first
supposed by Gurevich (1939). Then Ivanenko and Kurdgelaidze (1965)
and Itoh (1970) advanced hypotheses concerning stars composed of
quark matter. Nowadays there exists an extensive literature on
neutron stars, quark stars, and neutron stars containing quark
cores (the so-called hybrid stars). The excellent monograph of
 Haensel et al. (2007) can be recommended for further reading.

In a hybrid star the quark core is separated from the outer
nuclear matter envelope by an intermediate layer where the phase
transition (PT) between nuclear and quark matter occurs. Within
this region of coexistence of nuclear and quark phases there is
the possibility that the pressure is decreasing with growing
temperature at constant density. Such an effect was mentioned, for
instance, by Yudin et al. (2013), Hempel et al. (2013), and
Iosilevskiy (2014). In single phases of quark or hadronic matter
usually the opposite is the case, i.e., the pressure is increasing
with growing temperature at constant density. According to the
Clapeyron-Clausius formula (Landau and Lifshitz 1980) the
derivative of pressure over temperature $\DERPT{P}$ along the
phase transition line in the phase diagram is given by
\begin{equation}
\DERPT{P}=\frac{S_2-S_1}{\frac{1}{\rho_2}-\frac{1}{\rho_1}}<0.\label{Clayperon}
\end{equation}
Here $S_{1,2}$ and $\rho_{1,2}$ are the entropies per baryon and
densities of phases 1 and 2 in the region of their coexistence. We
consider here the simplest Maxwellian description of the PT. In
this case, $\DERPT{P}$ in Eq.~(\ref{Clayperon}) can be replaced by
$\left(\frac{\partial P}{\partial T}\right)_{\rho}$, as long as
$\rho$ corresponds to a value where one has phase coexistence. The
negative sign of the derivative in Eq.~(\ref{Clayperon}) thus
appears because the quark phase has  a higher entropy per baryon,
$S_2>S_1$ (note that $\rho_2>\rho_1$). This also implies that to
go from the low-density phase 1 to the high-density phase 2 at
fixed temperature the system absorbs a thermal energy per baryon
of $\triangle q=T(S_2-S_1)>0$.

\begin{figure}
\centering
\includegraphics[width=0.8\columnwidth]{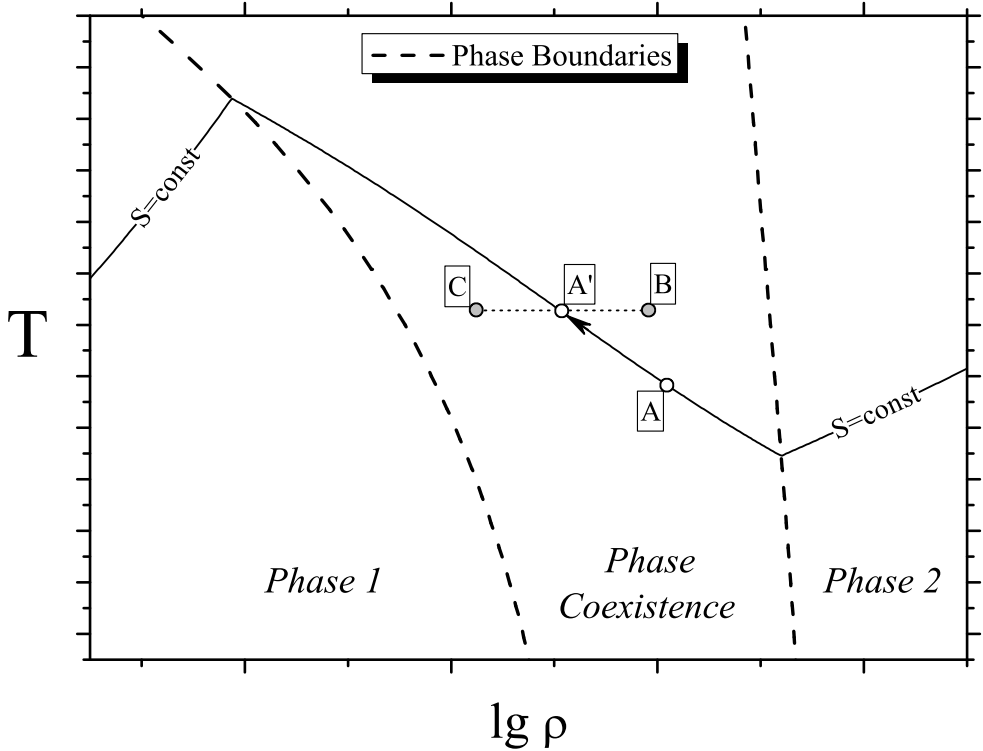}
\caption{Schematic view of a phase diagram for the transition
between hadronic and quark matter.} \label{Pix-PhaseDiagram}
\end{figure}

Let us consider a schematic view of the phase diagram in the
density-temperature plane (Fig.~\ref{Pix-PhaseDiagram}). The
boundaries of the phases are shown by bold dashed lines. The
region between the lines corresponds to two-phase
coexistence.\footnote{Often the region of phase coexistence is
also called the ``mixed phase''.}
An isentropic curve $(S=\mathrm{const})$ is shown by the solid
line. One can observe that in the phase-coexistence region the
temperature begins to decrease along the isentropic curve when the
density increases. Such a behavior of the PT was indicated by a
number of authors, and especially by Steiner et al. (2000). Note
that the temperate-pressure diagram would look qualitatively
similar as Fig.~\ref{Pix-PhaseDiagram}.

The main idea of our work is to investigate the properties of
convection expected in the region of phase coexistence. Consider
for example a convective element that starts its adiabatic motion
in the interior of a hybrid star at point $A$ and moves outwards
reaching point $A'$. So it keeps the initial entropy but has along
its way the same pressure and hence temperature (let us remind
that we consider Maxwellian PT) as the environment. If on its way
the convective element proves to be more dense than the
environment (point $C$) it will begin to sink. Such a
configuration is convectively stable. On the contrary, when the
state of the environment corresponds to point $B$, the density of
the convective element is lower than that of the environment and
it continues to rise --- the configuration is convectively
unstable. The entropy of environment in point $B$ is higher than
in point $A'$. Therefore the condition of appearance of convection
reads
\begin{equation}
\frac{dS}{dr}>0.
\end{equation}
This condition has the inequality sign opposite to the common
Schwarzschild criterion. Our paper is devoted to the consideration
of such uncommon circumstance and following consequences.

\section*{THE CONDITION OF APPEARANCE OF CONVECTION}
\noindent The criterion of appearance of convection in General
Relativity can be deduced in the same way as for Newtonian gravity
(Thorne 1966). One has only to take into account that the energy
per unit volume $\epsilon$ (with the rest mass energy being
included) plays the role of density. For the beginning of
convection the internal energy $\WID{\epsilon}{ce}$ of an
ascending convective element should remain less than that of the
surrounding matter $\WID{\epsilon}{sm}$ and vice versa for an
descending element:
$(\WID{\epsilon}{ce}-\WID{\epsilon}{sm})\triangle r<0$, where
$\triangle r$ is a radial displacement.

Taking into account that the convective element moves
adiabatically we can write down the following equations for
changes in internal energies $\WID{\epsilon}{ce}$ and
$\WID{\epsilon}{sm}$ on the way $\triangle r$:
\begin{equation}
\left\{
\begin{aligned}
\triangle\WID{\epsilon}{ce}&=\DERS{\epsilon}{P}{S,Y}\!\!\triangle P,\\
\triangle\WID{\epsilon}{sm}&=\DERS{\epsilon}{P}{S,Y}\!\!\triangle
P+\DERS{\epsilon}{S}{P,Y}\!\!\triangle
S+\DERS{\epsilon}{Y}{P,S}\!\!\triangle Y,
\end{aligned}
\right.
\end{equation}
where $P$ is the pressure, $S$ is the entropy per  baryon, and $Y$
is an additional dimensionless independent variable describing the
chemical composition of matter. For supernova matter the values of
the electron fraction $Y=\WID{Y}{e}$ or lepton charge fraction
$Y=\WID{Y}{e}+\WID{Y}{\nu}$ are usually specified. In the case of
several independent variables $Y_i$ describing the composition one
has to use sum over $i$ in the last term of the above expression
for $\triangle\WID{\epsilon}{sm}$. In the equation above we have
thus assumed that the convective element maintains its original
value of~$Y$.

Now we can write the criterion of appearance of convection in the
well known Ledoux form
\begin{equation}
\DERS{\epsilon}{S}{P,Y}\frac{dS}{dr}+\DERS{\epsilon}{Y}{P,S}
\frac{dY}{dr}>0.\label{Ledoux}
\end{equation}
For simplicity we assume below $Y=\textrm{const}$. Then the onset
of convection depends on the distribution of entropy in the star
(term $dS/dr$ in Eq. (\ref{Ledoux})) and the sign of the term
$\DERS{\epsilon}{S}{P,Y}$. In the nonrelativistic limit we have
$\epsilon\approx\rho c^2$ with $\rho$ being the baryon mass
density. Therefore up to a factor $c^2$ the multiplier in front of
$dS/dr$ is given by
\begin{equation}
\DERS{\rho}{S}{P}=-\frac{\rho^2}{\DERS{P}{T}{S}}=
-\frac{T\rho}{P\gamma\WID{c}{V}}\DERS{P}{T}{\rho},\label{drho_dS}
\end{equation}
where we introduce the adiabatic index $\gamma$ and the specific
heat capacity $\WID{c}{V}$:
\begin{equation}
\gamma\equiv\DERS{\ln P}{\ln\rho}{S},\quad \WID{c}{V}\equiv
T\DERS{S}{T}{\rho}.
\end{equation}
For matter under common conditions the right hand side of Eq.
(\ref{drho_dS}) is obviously strictly negative and we obtain the
criterion for onset of convection in the Schwarzschild form:
\begin{equation}
\frac{d S}{d r}<0.\label{Shvarcshild}
\end{equation}
Hence, a negative gradient of entropy causes the onset of
convection. However, within the phase-coexistence region the
derivatives $\DERS{P}{T}{S}$ and $\DERS{P}{T}{\rho}$ can be
negative and as a result the Schwarzschild criterion changes its
sign. The negative entropy gradient in this region ensures the
convective stability while the the positive one stimulates the
development of an convective instability.

Let us consider now the general case. One can easily show that the
factor in front of $dS/dr$ in Eq. (\ref{Ledoux}) is equal to
\begin{equation}
\DERS{\epsilon}{S}{P,Y}=\rho
T\left[1-\frac{\epsilon+P}{T\DERS{P}{T}{S}}\right].\label{de_ds}
\end{equation}
In the nonrelativistic limit the second term in square brackets is
much larger than 1 and equivalent to Eq.~(\ref{drho_dS}). If
$\DERS{P}{T}{S}<0$ the sign of this term as well as the criterion
of convection change again.

\section*{A REMARK ABOUT ISOTHERMS}
\noindent Let us make a slight digression to illustrate the
formulae of the previous section with an example of convection in
isothermal stars which are in the state of total thermal
equilibrium. In the limit of Newtonian gravity the temperature is
constant throughout the star, $T=\mathrm{const.}$, and we have
\begin{equation}
\frac{dS}{dr}=\DERS{S}{P}{T}\frac{dP}{dr}.
\end{equation}
Using Eq. (\ref{drho_dS}) we obtain
\begin{equation}
\DERS{\rho}{S}{P}\frac{dS}{dr}=\frac{T}{\rho^2\WID{c}{P}}
\DERS{\rho}{T}{P}^2\frac{dP}{dr}\leq 0,
\end{equation}
where $\WID{c}{P}=T\DERS{S}{T}{P}$ is the heat capacity at
constant pressure and the last inequality is valid because the
pressure never increases toward the surface of a star in
hydrostatic equilibrium. Thus we have the well known result that a
nonrelativistic isotherm is always convectively stable.

For the relativistic isotherms the condition of thermal
equilibrium is $T e^\nu=\mathrm{const.}$ (see Tolman 1969). Here
the function $\nu$ is connected with the time component of the
metric tensor ($e^{2\nu}=g_{00}$) and can be found from Einstein
equations:
\begin{equation}
\frac{d\nu}{dr}=-\frac{1}{P+\epsilon}\frac{dP}{dr}.
\end{equation}
Therefore for a relativistic isotherm the temperature falls by
going outwards in the star proportionally to $\frac{dP}{dr}$:
\begin{equation}
\frac{1}{T}\frac{dT}{dr}=\frac{1}{P+\epsilon}\frac{dP}{dr}.\label{gradT}
\end{equation}
In this case the convective stability is not obvious. Using Eq.
(\ref{gradT}) one can find the entropy gradient:
\begin{equation}
\frac{dS}{dr}=\DERS{S}{T}{P}\frac{dT}{dr}+\DERS{S}{P}{T}\frac{dP}{dr}=
\frac{\WID{c}{P}}{P+\epsilon}\left[1-\frac{\epsilon+P}{T\DERS{P}{T}{S}}\right]\frac{dP}{dr}.
\end{equation}
Then finally using also Eq. (\ref{de_ds}) we obtain
\begin{equation}
\DERS{\epsilon}{S}{P,Y}\!\frac{dS}{dr}=\frac{\rho
T\WID{c}{P}}{P+\epsilon}\left[1-\frac{\epsilon+P}
{T\DERS{P}{T}{S}}\right]^2\!\frac{dP}{dr}\leq 0.
\end{equation}
Now one can see that also the relativistic isotherm is always
convectively stable.

\section*{RESULTS OF CALCULATIONS}
\begin{figure}
\centering
\includegraphics[width=0.9\columnwidth]{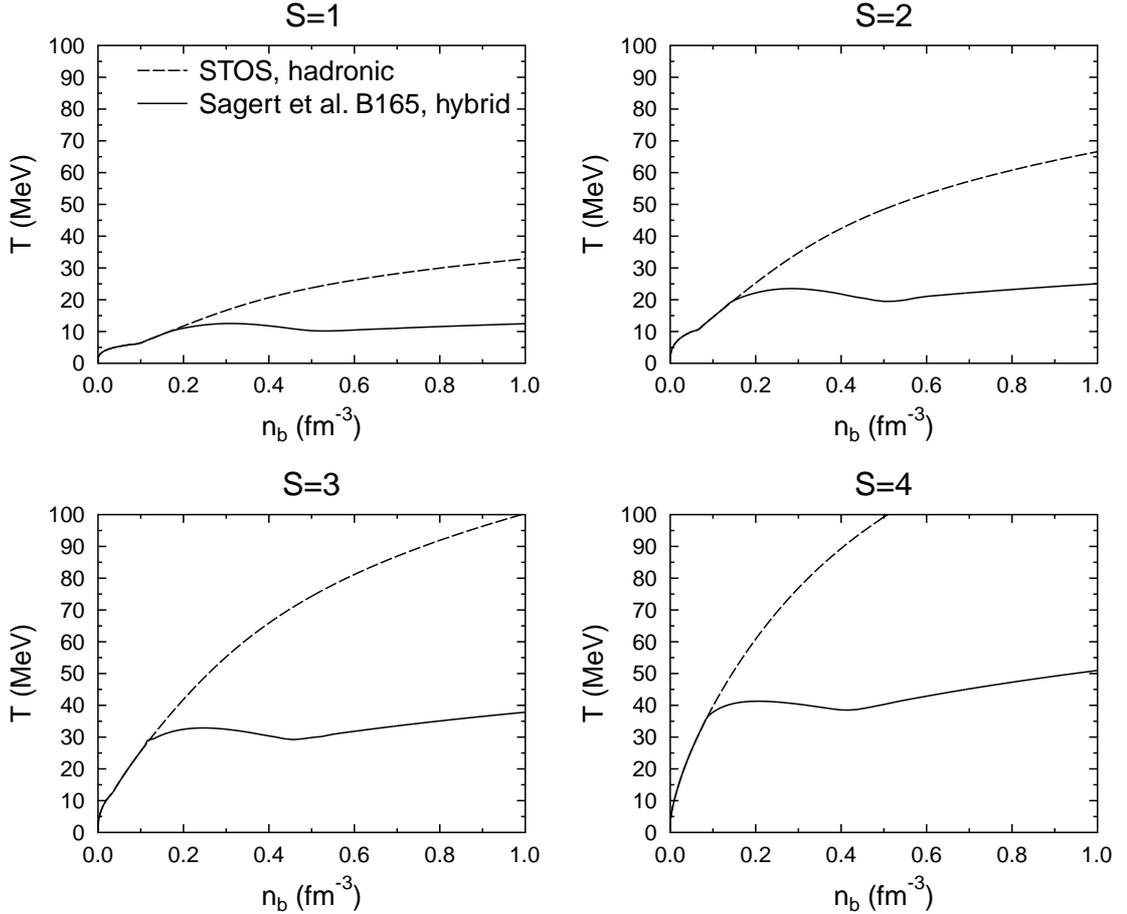}
\caption{The temperature versus baryon density $\WID{n}{b}$ for
hadronic (Shen et al. 2011) and hybrid equations of state (Sagert
et al. 2009) (dashed and solid curves, respectively) for various
different values of the entropy per baryon $S$ and a fixed lepton
fraction of $Y_L=0.4$.} \label{Pix-IsoEntrope}
\end{figure}

\begin{figure}
\centering
\includegraphics[width=0.9\columnwidth]{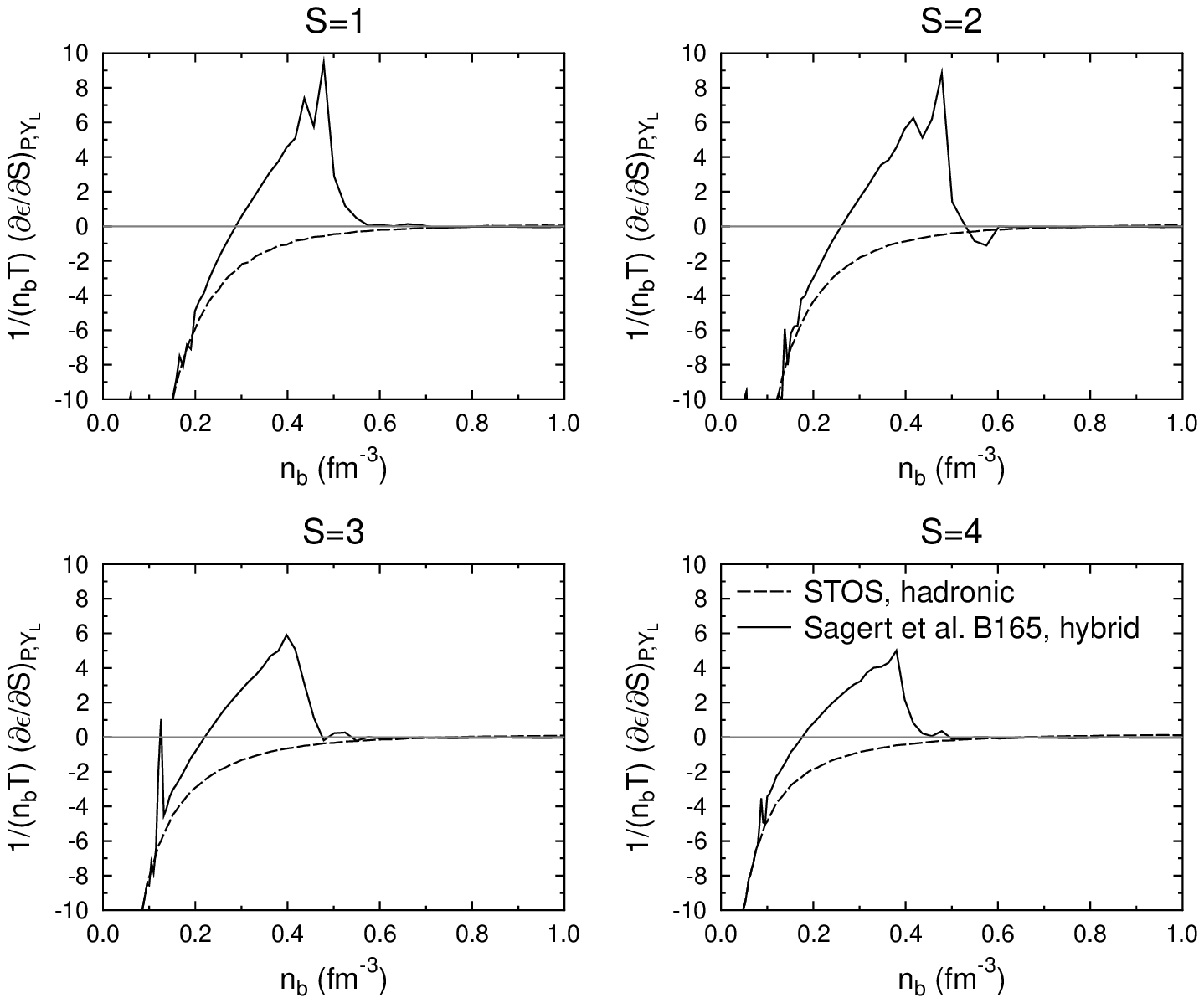}
\caption{Derivative $\DERS{\epsilon}{S}{P,Y}$ multiplied by
 $\frac{1}{n_b T}$ versus baryon
density $\WID{n}{b}$ for hadronic (Shen et al. 2011) and hybrid
equations of state (Sagert et al. 2009) (dashed and solid curves,
respectively) for various different values of the entropy per
baryon $S$ and a fixed lepton fraction of $Y_L=0.4$.}
\label{Pix-deds_b165}
\end{figure}

\noindent As an example of our calculations with an realistic
equation of state (EOS)
 we discuss the results shown in Figs. \ref{Pix-IsoEntrope} and \ref{Pix-deds_b165},
 where the dependencies of the temperature T and derivative $\DERS{\epsilon}{S}{P,Y}$
 on baryon density $\WID{n}{b}$ are shown for different values of
 the entropy per baryon  $S$.\footnote{We use natural units where $\hbar=c=k_B=1$.}
 The dashed curves in Figs. \ref{Pix-IsoEntrope} and \ref{Pix-deds_b165}
 correspond to the EOS for nuclear (hadronic) matter from Shen et al. (1998), Shen et al. (2011) (abbreviated STOS) whereas the
 solid ones refer to the hybrid EOS from Sagert et al. (2009), Fischer et al. (2011) which includes the
phase transition to quark matter
 using the quark bag model described by bag constant $B$.
 The quark bag constant $B^{1/4}$  and the lepton number per
 baryon $Y=\WID{Y}{L}=\WID{Y}{e}+\WID{Y}{\nu}$ are assumed to be $165~\mbox{MeV}$
 and 0.4,
 respectively.  The neutrino component is presumed to be in equilibrium with matter.
Global charge neutrality is assumed for phase coexistence, which,
as was first pointed out by Glendenning (1992), results in a Gibbs
or ``non-congruent'' (Hempel et al. 2013) phase transition.

The derivative shown in Fig.~\ref{Pix-deds_b165} corresponds to a
secondary thermodynamic derivative. We have calculated it from
thermodynamic consistent, tri-cubic interpolation of the free
energy. The used scheme is not able to handle properly the onset
and end of the phase transition, leading to the spikes visible in
Fig.~\ref{Pix-deds_b165} for the hybrid EOS that have to be
considered as numerical artifacts.

 The region of phase coexistence is distinctly identified with the negative temperature
 gradient in Fig. \ref{Pix-IsoEntrope} and positive sign of derivative
  $\DERS{\epsilon}{S}{P,Y}>0$ in Fig. \ref{Pix-deds_b165}. Exactly here the ``inverse''
  convection could be developed. One can observe such an effect also in fig.~6 from
  Fischer et al. (2011) where the results of calculations for different values
  of lepton number are presented.

 However, the situation is rather complex. First, along the same isentrope there are
 regions both with positive and negative sign of $\DERS{\epsilon}{S}{P,Y}$.
 Second, we have observed that the region of inverse convection is maximal for symmetrical matter with
 $\WID{Y}{e}=0.5$ and shrinks as $\WID{Y}{e}$ decreases when the neutronization
 is increasing. Therefore, the existence of the inverse convection and its parameters
 depend on:
\begin{enumerate}
 \item the equations of state that are assumed both for hadronic and quark matter;

 \item  the method of the PT description such as  Maxwellian, Gibbs (total thermodynamic
 equilibrium if Coulomb interactions are neglected), or as possible intermediate combinations (for a detailed
 discussion of various cases see Hempel et al. 2009). In
 reality, of course, this freedom of choosing the PT type is just because of our current lack of
 knowledge about important parameters of microphysical interaction among the phases such as surface tension
 (see e.g. Maruyama et al. 2007);

 \item the state of matter in the particular physical scenario considered,
 i.e., the values of the thermodynamic state parameters that are reached;
 first of all the entropy and the lepton number, and second the degree of matter transparency for neutrinos.
 \end{enumerate}

 In the near future we plan to investigate the above components that have an effect
 on the inverse convection and the possible influence of such convection on the
 subsequent evolution of hybrid stars.

\section*{DISCUSSION AND CONCLUSION}
\noindent First, we need to emphasize that the possibility of
``inverse''
 convection discussed here is the direct consequence of the
unusual property of the deconfinement PT
 expressed by Eq.~(\ref{Clayperon}). For example,
for the nuclear liquid-gas PT  the condition for convection will
have the ordinary form as for single phases of quark and hadronic
matter. For a comparison of these two PT, see, e.g.,~Hempel et al.
(2013).

 The consequences of possible existence of the
``inverse''
 convection zone in a hybrid star can be conceived by looking at
Fig.~\ref{Pix-PhaseDiagram}. In the case of well developed
convection an ascending matter element that travels from $A$ to
$A'$ has a lower entropy than the environment and thus a lower
heat content. Similarly, an descending element  has a higher heat
content than the environment. Therefore in contrast to ordinary
convection, within the ``inverse'' convection zone the heat flux
is directed inwards in a star.

Let us consider the situation when the new quark phase appears
inside a hot protoneutron star, formed shortly after the bounce of
the core of a collapsing massive star (see more details about the
collapse and its connection with supernova explosions in Sagert et
al. 2009 and Fischer et al. 2011). We consider that the conversion
of hadronic into quark matter proceeds on a timescale much faster
than the hydrodynamic evolution of the proto-neutron star (for
other scenarios, see the recent review by Drago and Pagliara 2015
and references therein). In this case, every matter element,
transformed from the hadron phase to the quark one keeps its
entropy per baryon $S$ and consequently decreases its temperature
(see our Fig.~\ref{Pix-PhaseDiagram} and also fig. 4 in Steiner et
al. 2000). Thus, the temperature of the hybrid star quark core
turns out to be lower than that of the surrounding hadronic
envelope. After the inverse convection has set in, the heat flux
is directed inwards, being transported by ``inverse'' convective
streams inside the phase coexistence domain. This could also be
accompanied by an inward neutrino flux which will try to remove
the temperature difference between the quark core and the
envelope, but this process requires special investigation.

After thermal relaxation the temperatures of the quark and hadron
phases become similar. However, the entropy of the quark phase
$\WID{S}{q}$ exceeds that of the hadron phase $\WID{S}{h}$. Hence,
within the region of phase coexistence we have the negative
entropy gradient that guarantees convective stability under the
condition of the ``inverse'' convection. After such internal
thermal relaxation the subsequent thermal evolution of the hybrid
star probably follows  the standard picture (see, e.g. Nakazato et
al. 2013). The detailed study of the thermal processes in question
is of considerable interest and may tell a lot about the birth and
thermal evolution of hybrid stars.

Recently Roberts et al. (2012) showed that different symmetry
energies of the nuclear EOS induce different strengths of (normal)
convection inside the cooling proto-neutron star after a
core-collapse supernova explosion. The corresponding neutrino
signal would allow to discriminate different hadronic EOS. It
would be very interesting to investigate the neutrino signal of
the ``inverse'' convection proposed here, and whether or not it
contains  a particular fingerprint of quark matter.

Currently, the neutrino-driven scenario is the favored explosion
mechanism of core-collapse supernovae. In this scenario turbulence
and convection are crucial to achieve sufficient neutrino heating
of matter to trigger an explosion. If the ``inverse'' convection
appeared already in the early post-bounce phase of the supernovae,
therefore in principle it could also impact the explosion
dynamics. These possibilities require further investigation by
detailed numerical simulations.

Possible direct effects of the pressure decrease in the region of
phase-coexistence on the stability of proto-neutron stars will be
the subject of a forthcoming study by some of the authors.

\vspace{1cm}
\section*{Acknowledgements}
The authors are thankful to the Swiss National Science Foundation
(SNSF) for support (grant SCOPES No.~IZ73Z0 152485). \ A.Y. thanks
also the RF Government grant No.~11.G34.31.0047. \ M.H. is
supported by the SNSF and partial support comes from NewCompStar,
COST Action MP1304. We also would like to thank the anonymous
referee for his/her constructive suggestions.



\begin{references}

\reference A.~Drago, G.~Pagliara, arXiv:1509.02134, (2015)

\reference T.~Fischer, I.~Sagert, G.~Pagliara, M.~Hempel,
J.~Schaffner-Bielich, T.~Rauscher, F.-K.~Thielemann, R.~Kappeli,
G.~Martinez-Pinedo, M.~Liebend\"{o}rfer, \apjs, \textbf{194}, 39,
28 pp., (2011)

\reference N. K.~Glendenning, Phys. Rev. D, 46, 4, pp. 1274-1287,
(1992)

\reference I. Gurevich, Nature, {\bf 144} 326-327, (1939)

\reference P. Haensel, A. Y. Potekhin, D. G. Yakovlev,
 {\em Neutron Stars 1 (Equation of State and Structure)\/},
 Springer, 619pp, (2007)

\reference M.~Hempel, V.~Dexheimer,
  S.~Schramm, I.~Iosilevskiy, \prc, {\bf 88}, 1, (2013)

\reference M.~Hempel, G.~Pagliara, and J.~Schaffner-Bielich, \prd,
\textbf{80}, 12, (2009)

\reference I.~Iosilevskiy, arXiv:1403.8053 [physics.plasm-ph],
(2014)

\reference N. Itoh, Prog. Theor. Phys., {\bf 44}, 291, (1970)

\reference  D. D. Ivanenko, D. F. Kurdgelaidze, Astrofizika, {\bf
1}, No. 4, 479–482, (1965)

\reference Landau L.D. and Lifshitz E.M., {\em Stastical
Physics\/},
 Part 1, 3d Edition, Oxford: Pergamon Press, (1980)

\reference T.~Maruyama, S.~Chiba, H-J.~Schulze, and T.~Tatsumi,
 Phys. Rev. D 76, 12, (2007)

\reference K.~Nakazato, K.~Sumiyoshi, H.~Suzuki, T.~Totani,
H.~Umeda, S.~Yamada, \apjs\ \textbf{205}, 1, (2013)

\reference L. F.~Roberts,  G.~Shen, V.~Cirigliano,  J. A.~Pons,
S.~Reddy, S. E.~Woosley,  Phys. Rev. Lett., 108, 061103, (2012)

\reference I. Sagert, T. Fischer, M. Hempel, G. Pagliara, J. Schaffner-Bielich,
           A. Mezzacappa, F.-K. Thielemann, M. Liebend\"{o}rfer,
           \prl, {\bf 102}, 081101, (2009)

\reference H. Shen, H. Toki, K. Oyamatsu, K. Sumiyoshi, Progr.
Theor. Phys., {\bf 100}, 1013-1031, (1998)


\reference H. Shen, H. Toki, K. Oyamatsu, K. Sumiyoshi, Astrophys.
J. Suppl., {\bf 197}, 20, (2011)

\reference A.~Steiner, M.~Prakash, J.~M.~Lattimer, \prb, {\bf
486}, 239-248, (2000)

\reference K.~Thorne, \apj, {\bf 144}, 201, (1966)

\reference R.C.~Tolman ``Relativity Thermodynamics and
Cosmology'', Oxford at the Clarendon Press, (1969)

\reference A.V.~Yudin, T.L.~Razinkova, D.K.~Nadyozhin
 Astronomy Letters, {\bf 39}, 161, (2013)

\end{references}
\end{document}